\documentclass[aps,superscriptaddress,prl,oneside,showpacs,amsmath,amssymb, reprint]{revtex4-1}
\usepackage{hyperref}
\usepackage{times}
\usepackage{graphicx}% Include figure files
\usepackage{dcolumn}% Align table columns on decimal point
\usepackage{bm}% bold math
\bibpunct{}{}{,}{s}{}{}

\usepackage{amsmath}    % need for subequations
\usepackage{graphicx}   % need for figures
\usepackage{verbatim}   % useful for program listings
\usepackage{color}      % use if color is used in text
\usepackage{braket}

\begin{document}

\title{Using coherent dynamics to quantify spin-coupling within triplet-exciton/polaron complexes in organic diodes}
\author{W. J. Baker}
\affiliation{School of Physics, UNSW, Sydney NSW 2052, Australia}
\affiliation{Centre for Quantum Computation and Communication Technology, UNSW, Sydney NSW 2052, Australia}
\affiliation{Department of Physics and Astronomy, University of Utah, Salt Lake City, UT 84112}
\author{T. L. Keevers}
\affiliation{School of Physics, UNSW, Sydney NSW 2052, Australia}
\author{C. Boehme}
\affiliation{Department of Physics and Astronomy, University of Utah, Salt Lake City, UT 84112}
\author{D. R. McCamey}
\email{dane.mccamey@unsw.edu.au}
\affiliation{School of Physics, UNSW, Sydney NSW 2052, Australia}

\date{\today}

\begin{abstract}
Quantifying the spin-spin interactions which influence electronic transitions in organic semiconductors is crucial for understanding their magneto-optoelectronic properties. By combining a theoretical model for three spin interactions in the coherent regime with pulsed electrically detected magnetic resonance experiments on MEH-PPV diodes, we quantify the spin-coupling within complexes comprising three spin-1/2 particles. We determine that these particles form triplet-exciton/polaron pairs, where the polaron--exciton exchange is over 5 orders of magnitude weaker ($<170$MHz) than that within the exciton. This approach providing a direct spectroscopic approach for distinguishing between coupling regimens, such as strongly bound trions, which have been proposed to occur in organic devices.
\end{abstract}

\maketitle

%%%%%%%%%%%%%%%%%%%%%%%%%%%%%%%%%%%%%%%%%%%%%%%%%%%%%%%%%%%%%%%%%%%%%%%%%%%%%%%%%%%%%%%%%%%%%%
%\section{intro and history}
%%%%%%%%%%%%%%%%%%%%%%%%%%%%%%%%%%%%%%%%%%%%%%%%%%%%%%%%%%%%%%%%%%%%%%%%%%%%%%%%%%%%%%%%%%%%%%%
Spin--spin interactions between charge carriers in organic semiconductors mediate spin--dependent electronic processes such as recombination and transport.  Quantifying these interactions is therefore crucial for understanding the macroscopic magneto--optoelectronic properties of these materials and devices made from them~\cite{Ando2013,Rao2013,Warner2013,Boehme2013}.Pulsed electrically detected magnetic resonance (pEDMR) allows the observation of coherent spin motion during the application of a magnetic resonant excitation of charge carrier spins, an experiment which provides a direct probe of interactions between those spin systems which control the sample conductivity. In this work, we focus for the first time on the application of pEDMR to  interactions involving more than two spins. In particular, we aim to quantify the coupling of the three spin-1/2 particles involved in the recombination process mediated by triplet exciton-polaron (TEP) interactions.\par

In recent years, the triplet exciton--polaron pair (TEP) model has been invoked in several studies to successfully explain magnetoresistance in organic semiconductors~\cite{Janssen2013,Desai2007,Cox2014}. However, due to the ambiguity of linking magneto-optoelectronic materials properties with their underlying spin--dependent processes, spectroscopic confirmation of any claimed microscopic process is needed~\cite{Boehme2013}. This is especially true for experimental conditions where other spin--states, such as strongly bound trions, and associated spin-dependent process can exist.

\begin{figure}[t!]
\includegraphics[width=8.5cm]{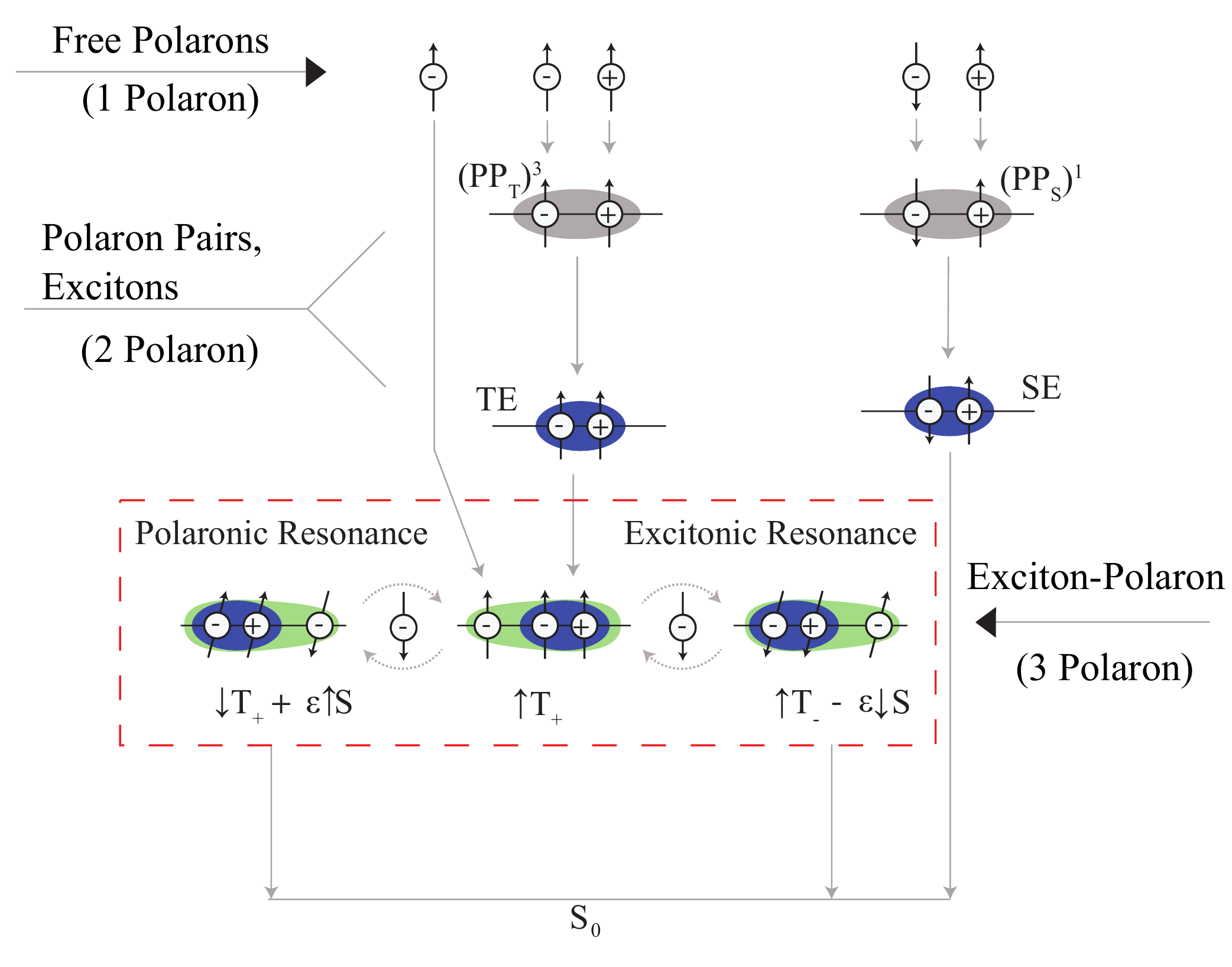}
\caption{Illustration of the TEP mechanism and the influences on conductivity upon ESR excitation. Coulombically coupled pairs of electrons and holes (so called polaron pairs, $PP_{i}$) randomly recombine into excitonic states. Thereafter, the long--lived triplet excitons (TE) can interact with polarons forming TEP pairs at the focus of this study. Due to the mixed eigenstates of TE and polarons via spin-coupling(see text) the singlet content and thus the TEP recombination and sample conductivity can change upon the application of ESR.}
\label{fig1}
\end{figure}

PEDMR is an ideal experimental technique with which to obtain such confirmation. It provides a more sensitive method for investigating thin film devices than conventional pEPR \cite{McCamey2006a}. More importantly, pEDMR directly discriminates between paramagnetic species which are involved in spin-dependent transitions and those which are not. We recently demonstrated that pEDMR could be used to observe the TEP process in experiments on $\pi$-conjugated polymer Poly[2-methoxy-5-(2-ethylhexyloxy)-1,4-phenylenevinylene] (MEH-PPV) based thin--film OLEDs.\cite{Baker2011}.The study presented in the following builds on this work by developing a theoretical framework for pEDMR on three spin complexes, then using that framework to analyse experimental results obtained on MEH-PPV OLEDs to quantitatively distinguish the TEP process from alternative coupling regimes (ie trions).

Figure \ref{fig1} depicts a hierarchy of spin configurations in an organic semiconductor under charge carrier injection, including TEP states. From the continuum of injected free charge carriers, weakly spin interacting polaron pairs form~\cite{Frankevich1992, McCamey2008}, which can either dissociate back to free charge carriers or recombine via the excitonic states. Due to the singlet nature of the ground state and the requirement of spin--conservation for electronic transitions, singlet excitons can recombine directly while the triplet excitons require further interaction with the environment which causes them to exhibit longer lifetimes. The TEP process requires triplet excitons to interact with excess charge carriers to create triplet--exciton polaron complexes (see Fig.\ref{fig1}). As proposed by Ern and Merrifield\cite{Ern1968}, the exciton relaxes to the ground state non-radiatively, transferring its energy excess to the (now free) polaron, changing the conductivity in the device. Since this Auger-like recombination transition is dependent on the spin state of the three $s=1/2$ manifold, it can be directly controlled with electron spin resonance (ESR), either via the polaronic or the excitonic resonance (Fig.\ref{fig1}), and detected by monitoring the conductivity of the sample.

\begin{figure}[t!]
\includegraphics[width=8cm]{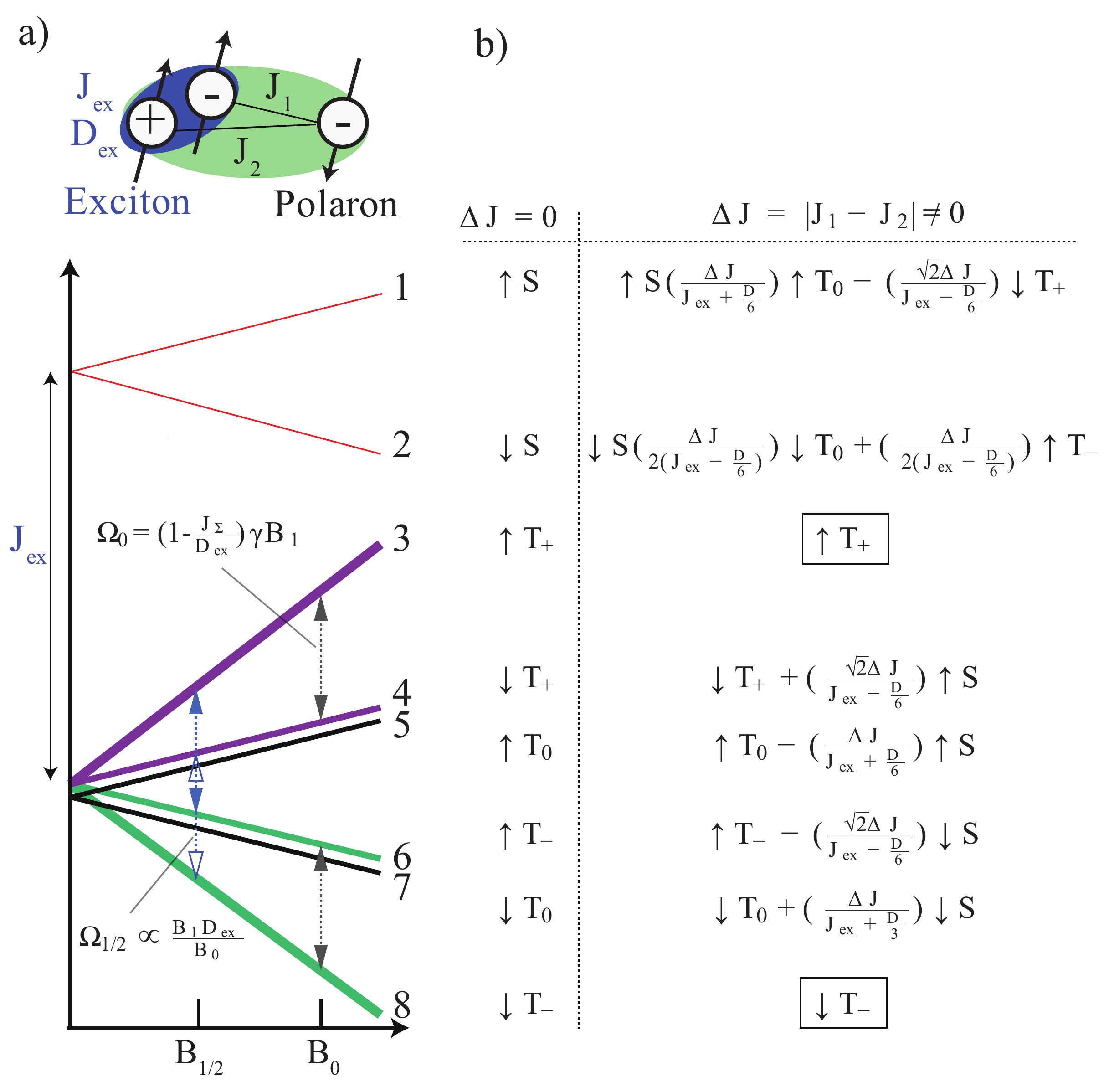}
\caption{Depiction of weakly spin--spin coupled exciton/polaron states with $\Delta J < J_{\Sigma} \ll J_{ex}, D_{ex} $). a) The energy term diagram of the eight spin eigenstates as a function of the applied external magnetic field $B_0$ with the arrows showing the ESR allowed transitions at full and half field conditions. Notice the separation in spin states at full-field governed by $J_{\Sigma}$ (see text).  b) Only when the small spin-coupling interaction between the exciton and polaron is non-vanishing ($\Delta J \neq 0$), a mixture of the exciton singlet and triplet--states occurs. The two states surrounded by boxes represent the dominantly occupied states under steady state device conditions before an ESR excitation takes place.}
\label{fig2}
\end{figure}

Figure~\ref{fig2}a) shows a depiction the eight spin eigenstates of a three $s=1/2$ complex with strong exchange coupling $J_{ex}$ and strong magnetic dipolar interaction $D_{ex}$ between two of the three electron spins as a function of an externally applied magnetic field. The two strongly coupled spins form an exciton state with $s=1$ which is then coupled through much weaker exchange coupling $J_{i}$ to the remaining electron spin. $J_{i}$ represents the individual exchange interaction between the free polaron and one of the carriers within the exciton. Our analysis shows that the frequency of nutation between the spin eigenstates under spin resonant excitation is governed by a sum of the two $J_{i}$ terms, $J_{\Sigma}$, and importantly, that a small difference in the $J_{i}$ terms $\Delta J$ is needed to provide an observable change in recombination rate. Details of this analysis can be found in Ref. \onlinecite{Keevers2015}, along with calculations of the energies of these spin eigenstates displayed in Fig.~\ref{fig2}a). At zero magnetic field, the energy scale is dominated by $J_{ex}$ leaving the singlet manifold by $\approx0.7$eV = 170 THz\cite{Reufer2005} higher in energy than that of the triplet exciton.

 When $\Delta J\neq 0$ (see Fig.\ref{fig2} b), right colum), the presence of the additional polaron causes a mixing of the triplet spin eigenstates, except for the $|\uparrow\rangle\otimes|T_+\rangle$ and $|\downarrow\rangle\otimes|T_-\rangle$ states. It is this mixture which causes the TEP recombination rate to depend on the occupation densities of spin--eigenstates and to therefore change when these densities are changed by magnetic resonant spin manipulation. Figure\ref{fig2} a) shows ESR--allowed spin transitions between eigenstates 3-4 and 6-8 in the presence of a magnetic field that will be referred to as the high--field resonance in the following, and between eigenstates 4-8 and 3-6 for what will be referred to as the low--field resonance. As the solutions for the eigenstates in terms of the TEP product states in Figure\ref{fig2}b) illustrates, as long as $\Delta J=0$, the excitation of ESR--allowed transitions does not changes the triplet content of the excited exciton states and thus, pEDMR signals do not exist\cite{Keevers2015}.

When $\Delta J\neq 0$ and sufficiently large and a constant bias is applied to an MEH-PPV device with imbalanced (=majority carrier) injection conditions, the TEP process will generate a surplus of $|\uparrow\rangle\otimes|T_{+}\rangle$ and $|\downarrow\rangle\otimes|T_{-}\rangle$ states (eigenstates 3 and 8) when the recombination rate approaches the steady state. This surplus is caused by the longer lifetimes of these states compared to all other eigenstates. The excitation of transitions 8-6 and 3-4 from the steady state will then increase the singlet exciton content of the TEP ensemble and thus, cause a measurable change of the recombination rate and the sample conductivity. A similar argument can be made for transitions 3-6 and 4-8 which can be excited under half--field conditions.

\begin{figure}
\includegraphics[width=8cm]{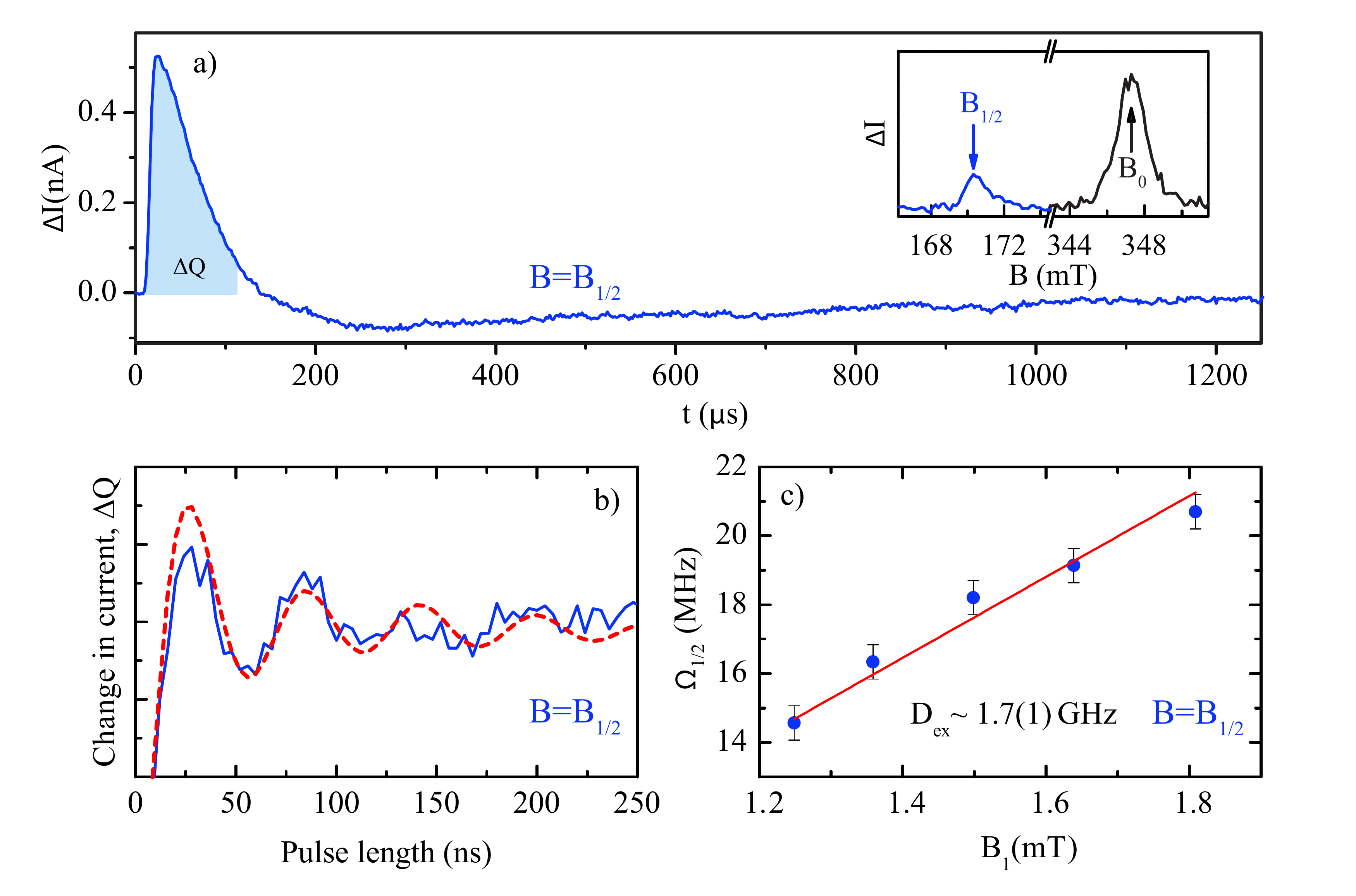}
\caption{(a) Plot of the device current $I(t)$ following a $\tau=200$ns microwave pulse at the resonance at $B_{\textrm{1/2}}\sim$170mT (The inset shows the full and half-field resonance peaks). The shaded blue area indicate the integration interval of the current measurements displayed in (b). (b) The solid line represents the integrated current change $Q$ as a function of the applied microwave pulse length recorded at the center of the half--field resonance. The dashed line shows a best fit simulated curve based on the TEP model. It reveals the Rabi--oscillation frequency $\Omega_{\textrm{1/2}}$ (c) The data points display Rabi--oscillation frequency $\Omega_{\textrm{1/2}}$ values obtained by fitting experimental data similar to the data set shown in (b) for various microwave field strengths $B_1$. The red line is a linear fit of the blue data points which shows good agreement. The slope of this linear fit reveals a value of the exciton's spin--dipolar coupling strengths $D_{ex}=1.7(1)$GHz.}
\label{fig3}
\end{figure}

For the pEDMR experiments discussed in the following, multilayer thin--film organic diode devices were fabricated consisting of an ITO/MEH-PPV/Ca/Al stack using a previously described device architecture\cite{McCamey2008, Baker2012,Baker2012a} for organic light emitting diodes excluding the hole injection layer used for bipolar injection~\cite{Baker2011}. In an externally magnetic field, $B$, we apply short bursts of intensive X-band microwave ($\approx 9.7$GHz) pulses to excite ESR--allowed transitions between the TEP spin eigenstates. The resulting changes of the device current $I\left(t\right)$ were recorded as a function of time. The inset of Fig.\ref{fig3}a) shows $I(t)$ following a $\tau=200$ns pulse as a function of the applied magnetic field $B$, similar to that seen in Ref.~\cite{Baker2011}. Two resonant changes in $I(t)$ are visible around magnetic field values of $B_{0}\approx 347$mT and $B_{\textrm{1/2}}\approx 170$mT (slightly less than half of $B_{0}$). The full--field signal, occurring at magnetic fields corresponding to a Land\'e--factor of $g\approx2.002$, is due to both the TEP and also the polaron-pair mechanisms~\cite{McCamey2008, Baker2012,Baker2012a}. In contrast to the full-field pEDMR signal, the half--field current response is solely governed by the TEP process as polaron-pairs do not contain pair partners with $s=1$. Figure 3a) displays the current change $\Delta I\left(t\right)$ after a microwave pulse at $ B_{\textrm{1/2}}$. It displays the double exponential dynamics, consistent with theoretical modeling \cite{Keevers2015} and qualitatively similar to the response seen in 2 spin-1/2 systems\cite{boehme2003}.

With the pEDMR detected pure TEP signal at half--field conditions shown in Fig.~\ref{fig3}a), we can now study the dynamics of coherent spin motion on a nano-second timescale by measuring the current transient as a function of the length of the applied microwave pulse ~\cite{boehme2003} as previously reported for $B_{\textrm{1/2}}$ by Ref.~\onlinecite{Baker2011}. The result of this experiment, shown by the blue data in Fig.~\ref{fig3}b), reveals a rapidly damped oscillatory behavior caused by the spin--Rabi oscillation of the triplet exciton states. We have developed a fit procedure for this data based on the simulation of the TEP system using a spin--Liouville equation based on a pair Hamiltonian for a s=1/2 and an s=1 spin. The simulation of the spin--Rabi oscillation controlled currents shown by the red data in Fig.~\ref{fig3}b) show good agreement with the experimental data when appropriate simulation parameters (e.g. the coupling strengths) are chosen.

Figure~\ref{fig3}b) shows that exciton-polaron complexes can be resonantly controlled with microwaves and this can be used to study the dynamics of coherent TEP spin--motion. The resonance signal observed at $B_{\textrm{1/2}}$ is predominantly due to triplet--exciton spin transitions with $\Delta m =\pm 1$ [transitions 3-6 and 4-8 in Fig.~\ref{fig2}a)]. These transitions have previously been studied with both EDMR and ODMR experiments~\cite{Kollmar1982,Swanson1992,Baker2011}, but only with continuous wave (cw) adiabatic field sweep experiments. The remainder of this work will focus on the information that can be extracted by examining these effects. 

The existence of the excitonic $\Delta m =\pm 1$ half--field resonance provides the first indication of a dipolar interaction within the exciton, as spin--dipolar interactions introduce off-diagonal elements that mix the $T_{-}$ and $T_{+}$ states. This produces a small but non-negligible ESR--transition probability for these otherwise forbidden transitions~\cite{Waals1959}. Due to the reduced magnitude of the static field, the commonly used rotating wave approximation is no longer valid~\cite{Slichter}. In order to account for this, a second order expansion of the time-averaged Hamiltonian was used for the simulation. This revealed a half-field Rabi oscillation frequency $\Omega_{\textrm{1/2}}\propto \frac{B_{1}D_{ex}}{B_{0}}\sin\left(2\theta\right)$ in which the angle $\theta$ represents the orientation of the laboratory frame (governed by the external magnetic field orientation) with regard to the molecular frame of the triplet state\cite{Keevers2015}. The measured TEP spin--Rabi oscillation frequency can therefore be used to quantify the exciton dipolar interaction strength $D_{ex}$. In order to do this, we have repeated the measurements of the resonantly induced spin--Rabi oscillation at half--field conditions for various driving fields $B_1$ \footnote{$B_1$ was calibrated by measuring the Rabi oscillation frequency of the polaron-pair signal at full field.}. The results of these measurements were fit in the same way as the data shown in Fig.~\ref{fig3}b). The nutation frequencies $\Omega_{\textrm{1/2}}$ obtained from this procedure are plotted in Fig.~\ref{fig3}c) as a function of $B_1$. This plot shows a good agreement with a linear fit function [red line in Fig.~\ref{fig3}c)] and from its slope, along with the known applied static and applied oscillating magnetic fields, we obtain a dipolar interaction strength within the exciton of $D_{ex}=1.7(1)$GHz. This result matches previously determined values for the exciton zero-field parameters in PPV blends\cite{Swanson1990}, providing support for the model. 

\begin{figure}[t!]
\includegraphics[width=8cm]{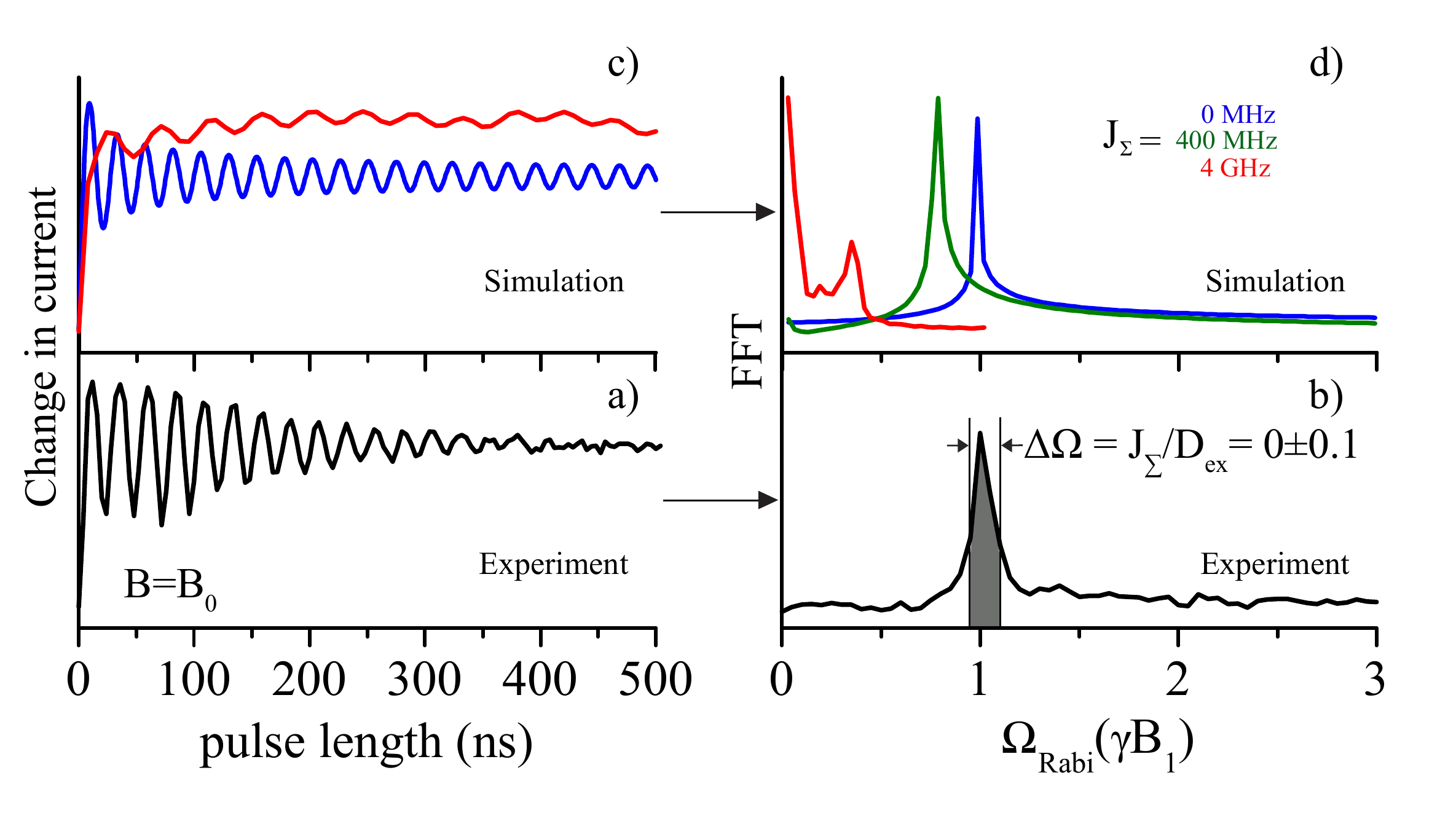}
\caption{a) Plot of the integrated current change $Q$ as a function of the applied microwave pulse length recorded at the center of the full--field resonance. (b) Fourier transform of the experimental data displayed in (a) with a frequency scale in units of $\gamma B_1$ with $\gamma$ being the gyromagnetic ratio. (c) Simulated values of the integrated current based on the TEP model for values of the $J_{\Sigma}$ corresponding to 0 MHz (blue), 400MHz (green), and 4GHz (red). (d) Fourier transform of the simulated data displayed in (c) with a frequency scale in units of $\gamma B_1$. The comparison of the experimentally obtained spin--Rabi frequency distribution with the simulated data sets shows that best agreement is obtained for $J_{\Sigma}= 0 \pm 170 $MHz. The error, based on the FWHM of the FFT data shown in panel d, sets an upper bound on $J_{\Sigma}$.}
\label{fig4}
\end{figure}

We now utilise this model to determine the intrapair--exchange between the triplet exciton and the polaron from the full--field spin--Rabi oscillation measurement. Figure \ref{fig4} a) (black curve) displays the spin--Rabi oscillation reflected by the device current after the application of a resonant microwave pulse at $B_{0}$ as a function of pulse duration, which corresponds to the rotation of the polaron s=1/2 particle within the complex with a Rabi nutation frequency of $\gamma_e B_1$, where $B_1$ is the magnitude of the oscillating field and $\gamma_e$ the electron polaron's gyromagnetic ratio. Neglecting states with large steady-state singlet content, the observed signal is caused by transitions 4 to 3  and 7 to 8 in Fig.\ref{fig2}a), leading to an increase in polaron mobility and thus, an increase in current. The spin--Rabi oscillation seen in Fig.~\ref{fig4}a) was recorded at $B_{0}$ with the same microwave frequency as the data in Fig.~\ref{fig3}b). While the oscillation at the half--field condition is caused by spin--Rabi oscillation of the exciton, the observation at the full--field condition is due to the polaron spin--Rabi oscillation.

%%%%%%%%%%%%%%%%%%%%%%%%%%%%%%%%%%%%%%%%%%%%%%%%%%%%%%%%%%%%%%%%%%%%%%%%%%%%%%%%%%%%%%%%%%%%%%%%%
%\section{description of model, what we did to understand the data}
%%%%%%%%%%%%%%%%%%%%%%%%%%%%%%%%%%%%%%%%%%%%%%%%%%%%%%%%%%%%%%%%%%%%%%%%%%%%%%%%%%%%%%%%%%%%%%%%%

The TEP model explains the occurrence and frequency of the oscillation of both the full- and half--field pEDMR signals. At full field, the excitation is in resonance with both polaron and excitonic transition as shown in Fig. \ref{fig3} (transitions 6-8 and 4-3 as well as 4-7 and 5-6). However, transitions 6-8 and 4-3 will dominate the signal due to the strongly quartet-like character of the steady state. In order to fully account for all environmental variations, we repeated the calculation for the full range of anisotropic dipolar and random hyperfine interactions and find for all cases that the polaron signal dominates at full--field.

As we adjust the interaction strengths between the exciton and polaron, the energy levels of the corresponding energy eigenstates are also modulated by the sum of the spin-coupling between the polaron and exciton entities $J_{\Sigma} = J_{1} + J_{2}$, leading to a change in the expected resonant frequency of the observed Rabi oscillation at full-field, $\Delta\Omega = -\gamma B_{1} J_{\Sigma}/D_{ex}$. This dependence of the full--field Rabi--frequency on $J_{\Sigma}$ along with the dipolar parameter obtained from the half-field resonance discussed above can be used to quantify the spin-coupling strength between the exciton and polaron. Figure \ref{fig4}(c) shows simulation results of the expected current--detected full--field spin--Rabi oscillation for various values of $J_{\Sigma}$. Figure \ref{fig4}(d) displays the frequency components of these transients in a Fourier transformation of the data in Fig.\ref{fig4}(c). The comparison with the Fourier transform of the experimental as shown in Fig.~\ref{fig4}(b) shows agreement with simulations where $J_{\Sigma} \leq 170$MHz. Thus, an upper bound of 0.7$\mu$ev can be placed for the triplet--exciton polaron spin-coupling strength, a value that is 5 to 6 orders of magnitude weaker than the excitonic exchange. 

Since the anisotropic nature of the dipolar coupling effect would cause a broadening of the Rabi frequency distribution \cite{Lee2010}, an upper bound of less than 5MHz based on the width of the peak in Fig.~\ref{fig4}(b) can be determined. The TEP intrapair dipolar interaction is therefore insignificant compared to the exchange interaction.

In conclusion, we report on a study comparing magnetic resonantly induced characteristic spin--motion signatures of TEPs on the device current on MEH-PPV diodes with calculations based on the TEP model. This procedure confirms the previously reported spin--spin coupling strengths within the triplet exciton states and reveals very weak exciton--polaron coupling.
Since $J_{\Sigma}\leq 170$MHz, we conclude that TEPs are weakly spin--exchange interacting exciton-polaron complexes while trion states, which have been invoked previously~\cite{Cox2013}, do not significantly contribute to the observed conductivity effects. This could be due to a short lifetime of the trion states or due to the circumstance that spin-coupling strength prevents an ESR allowed transition from affecting charge carrier conductivity.

%%%%%%%%%%%%%%%%%%%%%%%%%%%%%%%%%%%%%%%%%%%%%%%%%%%%%%%%%%%%%%%%%%%%%%%%%%%%%%%%%%%%%%%%%%%%%%%%%
%\section{summerizing the importance of the study and observation...sell it! bounds}
%%%%%%%%%%%%%%%%%%%%%%%%%%%%%%%%%%%%%%%%%%%%%%%%%%%%%%%%%%%%%%%%%%%%%%%%%%%%%%%%%%%%%%%%%%%%%%%%%

  \acknowledgments
  We acknowledge support from the Australian Research Council (DP12102888). DRM is supported by an ARC Future Fellowship (FT130100214) and TK by the Australian Renewable Energy Agency. The experimental data presented in this study was measured with support by the U.S. Department of Energy, Office of Basic Energy Sciences, Division of Material Sciences and Engineering under Award DE-SC0000909.

\bibliographystyle{apsrev4-1}

\end{document}